\documentclass[12pt,preprint]{aastex}

 \shorttitle{SAO 40039}
 \shortauthors{Sumangala et al.}

\begin{document}

 \title{IS THE POST-AGB STAR SAO\,40039 MILDLY HYDROGEN-DEFICIENT$?$}
 \author{S. Sumangala Rao\altaffilmark{2}, Gajendra Pandey\altaffilmark{2}, David L. Lambert\altaffilmark{3} \and Sunetra Giridhar\altaffilmark{2}}
 \altaffiltext{2}{Indian Institute of Astrophysics, Bengaluru-560034, India;\\
 sumangala@iiap.res.in; pandey@iiap.res.in; giridhar@iiap.res.in}
 \altaffiltext{3}{W. J. McDonald Observatory, The University of Texas, 
  Austin, Texas, USA 78712;\\ dll@astro.as.utexas.edu}

 \begin{abstract}

 We have conducted an LTE abundance analysis for SAO\,40039,
 a warm post-AGB star whose spectrum is known to show surprisingly strong
 He\,{\sc i} lines for its effective temperature and has been suspected of being H-deficient and He-rich.
 High-resolution optical spectra are analyzed using a family of model
 atmospheres
 with different He/H ratios. Atmospheric parameters are
 estimated from the ionization equilibrium set by neutral and singly
 ionized species of
 Fe and Mg, the excitation of  Fe\,{\sc i} and Fe\,{\sc
 ii} lines, and the wings of the Paschen lines.
 On the assumption that the He\,{\sc i} lines are of photospheric and
 not chromospheric origin, a
  He/H ratio of approximately unity is found by imposing the condition that
 the adopted He/H ratio
 of the model atmosphere must equal the ratio derived from the observed
 He\,{\sc i} triplet lines at 5876, 4471 and 4713\AA, and singlet
lines at 4922 and 5015\AA. 
 Using the model with the best-fitting atmospheric parameters for this He/H
 ratio,
 SAO\,40039 is confirmed to exhibit mild dust-gas depletion, i.e., the star 
has an atmosphere deficient in elements of high condensation temperature.
The star appears to be moderately metal-deficient with [Fe/H]=$-$0.4 dex. 
But the star's intrinsic metallicty as estimated from Na, S and Zn, elements of 
a low condensation temperature, is $\rm [Fe/H]_o$\footnote{$\rm [Fe/H]_o$ refers to the star's intrinsic metallicity}=$\simeq -0.2$.
The star is enriched in N and perhaps O too, changes reflecting the
star's AGB past and the event that led to He enrichment.

 \end{abstract}


 \section{Introduction}
 SAO\,40039, also known as IRAS 05040+4820 and BD $+48^\circ 1220$, is
 judged by
 several criteria to be
  a post-AGB star:
  its spectral type (A4Ia), its position in the IRAS color-color diagram
   and the double-peaked spectral energy distribution indicating a detached
 youthful cold dust shell
 around the central star. Using  VBRIJHK photometry and adopting a simple
 dust shell model,
  \citet{Fujii2002} find a
  dynamical age of 2460 years for the dust shell with dust at 97K, and
 estimate
 the star to have a core mass of
    0.55M$_{\odot}$ indicating  a low-mass
  progenitor for this PAGB star.

   \citet{Kloch2007} have presented
  an  abundance analysis  of SAO\,40039 based on  high-resolution spectra with a
 wavelength coverage of 4500-6760\AA. They have also discussed the star's
 radial
  velocity changes
  and also time-dependent differential velocity shifts between different
 classes of
  absorption lines. Their spectra showed a variable H$\alpha$ line
  with two emission components.
 Variable emission was also reported
 in H$\beta$ and in some lines of Si\,{\sc ii}, Fe\,{\sc i} and Fe\,{\sc
 ii}.
 But, in particular, these authors pointed out the abnormal strength of
 the
  He\,{\sc i} 5876\AA\ absorption feature in this star with the `low'
 effective
 temperature of about 8000 K.
 Their abundance analyses conducted with model atmospheres
computed for a normal He/H ratio (=0.1) gave the He/H ratio of
 0.7.

 Detection of He\,{\sc i} lines in absorption in several A and F-type PAGB stars
 has been
 reported previously. The 4471\AA\ line was found by \citet{Waelk92} in HD
 44179,
 the central star of the Red Rectangle.
  HD 187885, a PAGB star with  a high C/O ratio and an enrichment
 of $s$-process elements, shows the He\,{\sc i} 5876 \AA\ line \citep{Van96}.
 \citet{Klo2002} report five He\,{\sc i} lines in HD 331319. Detection of
 He\,{\sc i} lines in A and F-type supergiants suggest either a He
 enrichment, if the lines
 are of photospheric origin, or a contribution from a thick hot
 chromosphere of presumably a normal He abundance.
 The above studies adopt the assumption
 of a photospheric
 origin and report the photosphere to be moderately H-poor and He-rich.
 However, these analyses use model atmospheres computed for a
 normal He/H ratio and do not iterate on the construction of
 the model atmosphere and the  abundance
 analysis to obtain agreement between the input and output He/H ratio.

 Our goal in making a fresh analysis of SAO 40039 is to iterate to find the
 self-consistent He/H ratio. In undertaking the analysis, we  also adopt the
 assumption that the He\,{\sc i} lines originate in a
 He-rich photosphere and are not chromospheric in origin.

 \section{Observations and data reduction}

 Spectra of SAO\,40039 were obtained on the nights of 2007 November 5  and
 2011 February 21
 at the W.J. McDonald Observatory
  with the 2.7m
  Harlan J. Smith telescope and the Tull coud\'e spectrograph
 \citep{Tull95}. The spectra taken at these epochs were found to be similar.
  The spectra correspond to a resolving power of 60,000 (5 kms s$^{-1}$).
   Reduction of the raw data was performed
  with the Image Reduction and Analysis Facility (IRAF\footnote{The IRAF
 software is
 distributed by the National Optical Astronomy Observatories under contract
 with the National Science Foundation}) software package. 

  An additional spectrum was obtained on 2011 January 28 using the echelle
 spectrometer
  at Vainu Bappu Observatory in Kavalur, India
 giving a resolution of about 28,000 (10 kms s$^{-1}$) in slitless mode \citep{Rao2005}.
{\bf Since the projected rotational velocity of the star is about 15 kms s$^{-1}$, the lower resolution of the VBO spectrum does not affect the stellar absorption profiles and hence our measured equivalent widths and the calculated abundances.}

 \section{Abundance Analysis}

 Our spectra resemble closely those described by \citet{Kloch2007} with
 regards to
 emission in Balmer lines, the complex structure of the Na D lines and
 variable
 profiles of the metallic lines, and the  equivalent widths of lines
 unaffected
 by obvious emission.
 For our analysis we have used only clean,
  unblended and symmetric absorption lines. 

 We  have compared the equivalent widths of lines common between our spectra and 
those of \citet{Kloch2007}. 
A small systematic difference possibly caused by differences in resolution is 
discernible, but it appears that there were 
no major variations in the atmospheric parameters at these epochs. 
This is also supported by the fact that the average radial velocity of lines 
in our spectra (-12 kms s$^{-1}$) lies within the range as reported by
\citet{Kloch2007} 
in their spectra (-7 to -15 kms s$^{-1}$).

 Abundance analysis was done first on the assumption that the atmosphere has
 a normal He/H ratio.
 Model atmospheres for normal He/H ratio (He/H=0.1) were taken from
 the Kurucz database\footnote{http://kurucz.harvard.edu/grids.html}.
 Atmospheres of different He/H ratios were
  computed by the code STERNE \citep{jeff2001}.
   The LTE spectrum synthesis code MOOG (2009 version) by \citet{Sned74}
 was
 used with
 the grid of model atmospheres taken from the Kurucz database as mentioned above.
 The LTE code SPECTRUM \citep{jeff2001} was used with the STERNE
models.\footnote{Note that the
 stellar parameters and the abundances derived by adopting the 
 Kurucz
  models and the code MOOG for He/H = 0.1 are in excellent agreement with
 results
 from the STERNE models
  and the LTE code SPECTRUM, also for He/H = 0.1.}

 The procedure begins with the determination of the atmospheric parameters
 from
 the spectrum, continues with the abundance analysis for He and other
 elements and ends when the model of a certain He/H ratio reproduces the
 observed He\,{\sc i} lines.

 \subsection{Method for determining the stellar parameters}

 First, we estimated the microturbulent velocity ($\xi_{t}$) using Fe\,{\sc
 ii} lines with small range in 
 lower excitation potentials (LEP)(2.6$-$3.2 eV) as these lines have a good
 range
  in their equivalent widths thereby reducing any
 temperature
 dependence of the
 estimated $\xi_{t}$.
  The $\xi_{t}$ is found from the standard
  requirement that the abundance be independent
 of the measured equivalent width.
 The effective temperature ({\itshape T$_{\rm eff}$}) was estimated by the
 requirement that the abundance
 of a given species be
 independent of a line's LEP. This step was
 conducted independently for  Fe\,{\sc i} and
 Fe\,{\sc ii} lines as they are both well represented in the spectrum and
 show a range in their LEP's.
 Finally, the surface gravity ({\itshape $\log g$}) was estimated by the
 condition that there
 be ionization balance between the
 neutral and the singly ionized  Fe lines. This condition defines a locus
 in
 the ({\itshape T$_{\rm eff}$, $\log g$}) plane and the {\itshape T$_{\rm eff}$} 
 derived from Fe\,{\sc i} and Fe\,{\sc ii} breaks the degeneracy.
 This exercise is repeated for grids of model atmospheres with He:H ratios
 of 10:90, 30:70 and 50:50.

 A  microturbulence  $\xi_t = 4.8\pm1.0$ km s$^{-1}$ is found for all
 models
 with acceptable  effective temperature  and surface gravity and the value
 is
 insensitive to the He/H ratio of the models.
 The effective temperature {\itshape T$_{\rm eff}$} is found to be 8000$\pm300$K from
 26 Fe\,{\sc i} and
 20 Fe\,{\sc ii} lines. Within the 300K uncertainty, the temperature is
 independent of surface gravity over a considerable range and is not
 sensitive to the He/H ratio of the model atmosphere. 

 The {\itshape T$_{\rm eff}$}, {\itshape $\log g$} locus
 found from the ionization balance for Fe is illustrated
 in Figure 1 
 for the model atmosphere grid with He:H = 50:50. With {\itshape T$_{\rm eff}$} = 8000$\pm$300
 K from the excitation of Fe\,{\sc i} and Fe\,{\sc ii}, the
 {\itshape $\log g$}=0.75$\pm$0.25 cgs is found. 
  The surface gravity
 changes slightly with He:H running from 0.94 for He:H=10:90 through 0.83
 for
 He:H=30:70 to the above result for He:H=50:50. A second locus is provided
 by 
  Mg\,{\sc i} and 
 Mg\,{\sc ii} lines  -- see upper panel of Figure 1. The Mg and Fe loci are in good
 agreement. A third locus is offered by the Paschen lines (see below).

 \subsection{The He/H ratio}

Our spectra confirm the presence of the D3 5876\AA\ He\,{\sc i} line with an
equivalent width very similar to that reported by \citet{Kloch2007}.
 The strength of this line suggests that several
 other
 lines should be present unless the excitation of the D3 line is highly
 peculiar.
Our search also provided the additional triplets (4471 and 4713\AA) and two singlet lines
(4921 and 5015\AA).  Since the He\,{\sc i} at 4713\AA\ is blended with the Fe\,{\sc ii} line,
 its contribution has been included 
  while synthesizing the He\,{\sc i} profile at 4713\AA\ as  shown in the Figure 2, we get a consistent He/H ratio of approximately unity.     
Other possible He\,{\sc i} lines are predicted to be below the
detection limit, or blended with other lines or inaccessible due to inter-order gaps. 

 The He\,{\sc i} lines were computed for model atmospheres with
 ({\itshape T$_{\rm eff}$, $\log g$}, $\xi_{t}$)=(8000, 0.75, 4.8) and with
 the
 appropriate model for the
 He/H ratios: 0.1 (normal; solar abundance), 0.4 (30/70), and 1.0 (50/50).
 The synthesized profiles are convolved with the instrumental and the
 stellar rotation profiles,
 before matching with the observed spectrum as described in \citet{Pa2006}. 
The data for computing the He\,{\sc i} profiles were taken from \citet{jeff2001}.
 A projected rotational velocity of about 13$-$16 km s$^{-1}$ is
 estimated by using unblended moderately strong lines.
 As shown in Figure 2, the best fits to the observed He\,{\sc i} profiles
 at 5875, 4471, 4713, 4922 and 5015\AA\ were
 obtained for He/H ratio of 1.0 (50/50) with an uncertainty of
 approximately $\pm0.2$. 

  To verify the  model atmosphere parameters, we explore fits to the
 Paschen lines. The
 Balmer lines are  affected by emission and blends with other lines for
 their
 line profiles to be useful as monitors of the atmospheric parameters but the
 Paschen lines 
appear free of emission and blends. The Stark broadening tables for the Paschen
lines of hydrogen are taken from \citet{Lemke97}.
 In the lower panel of the Figure 1, we show
 observed and synthesized
 Paschen profiles near 8600\AA\ for the model
 with He/H=1.0  and
 {\itshape T$_{\rm eff}$} = 8000K and for surface gravities {\itshape
 $\log g$} = 0.5, 0.75 and 1.0.
 The best fitting profile is for {\itshape $\log g$} of 0.75. We also synthesized  the Paschen lines in the 8400\AA\ region for the above mentioned temperature and gravities and we get consistent results.
 These profiles are also
 sensitive to the adopted effective temperature with a 1000K increase
 approximately
 mimicking a 0.5 dex increase in surface gravity. The Paschen lines give the
locus shown in the upper panel of Figure 1.
 As the Paschen limit is approached, the Paschen lines overlap and depress
 the local
 continuum.  In addition, the location of the continuum is rendered more
 challenging
 because of the  echelle's blaze function. Yet, the synthesized profiles
 closer to the
 limit are able to reproduce the observed profiles quite well -- see lower panel of Figure
  1. 

 Our demonstration that SAO 40039 is H-deficient and He-rich rests on the
 assumption that
 the He\,{\sc i} absorption lines originate in the stellar photosphere and
 are not a
 product of a stellar chromosphere. (A second assumption is that the
 real stellar
 photosphere approximates the model atmospheres computed by classical
 procedures).
 A chromospheric origin is not trivially dismissed. One recalls that the D3
 line appears in
 absorption in spectra of F dwarfs with active chromospheres
 \citep{Wolff84,Danks1985};
  the correlation between D3 equivalent width and
 X-ray
 luminosity points strongly to a chromospheric origin. A chromospheric
 origin for
 He absorption is required to account for the presence of He\,{\sc i} 10830
 \AA\
 in spectra of warm-cool giants and supergiants where it may appear in
 absorption and/or
 emission \citep{O'Brien86} and cannot be of
 photospheric origin.

The possibility of the photospheric origin of the He\,{\sc i} lines in our spectrum 
is supported by several facts.
Firstly, the lines have the photospheric velocity.  Secondly and more importantly, 
the equivalent widths of all He\,{\sc i} lines -- singlets and triplets -- are
reproduced by a photospheric model. If the lines originated in the chromosphere, one
would expect a difference between singlets and triplets, as previously noted
by Klochkova et al. (2002) in their
discussion of the He\,{\sc i} absorption lines  found
 in HD 331319,
 a star similar to SAO 40039. 
 While our analysis do not preclude a chromospheric
 origin, we agree with
 Klochkova et al. who remark `the hypothesis of a photospheric origin for the
 He\,{\sc i}
 lines in HD 331319 is not rejected'.  
 Certainly, there must be a suspicion that SAO 40039's atmosphere is mildly
 H-deficient
 such that He/H= 1 now, a value indicating appreciable
 loss of H and enrichment of He since the star was a main sequence star
 with
 He/H $\simeq 0.1$.

 \section{Discussion and Results}

 Using the adopted model parameters ({\itshape T$_{\rm eff}$, $\log g$},
 $\xi_{t}$)=(8000, 0.75, 4.8) with
 He/H of 1.0, the measured equivalent widths of different species were used
 to derive the abundances. In Table 1 we present the abundances derived
 with the SPECTRUM code using the He/H=1.0 model and the abundances generated with the code MOOG using the He/H=0.1 model. 

A detailed line list (see for SAMPLE Table 2, that lists some lines of SAO 40039) used in our analysis has been presented electronically. The line list provides the lower excitation potential ($\chi$) for each line
, the {\itshape $\log gf$} value, sources of {\itshape $\log gf$}, the measured equivalent widths ($W_\lambda$) in m\AA\,
 the abundance (log $\epsilon$) derived from each line for the adopted model atmosphere.

 The solar abundances
for all the elements have been taken from \citet{Asplund2009}.

 SAO 40039 seems to be  mildly iron-poor. 
Klochkova et al.'s
 analysis assuming He/H= 0.1 gave a solar Fe abundance, a value essentially
 confirmed
 by our analysis for the model atmosphere also with He/H = 0.1. A principal
 effect of the
 higher He abundance is to reduce the continuous opacity per gram and so
 demand a
 lower iron abundance to match the same observed line strength.
 Inspection of the [X/Fe] entries in Table 1 points to several anomalies
 when the composition of
 an unevolved  disk  star is taken as the reference. The two most extreme
 anomalies are nitrogen  which is
 very overabundant ([N/Fe] = +1.1), and aluminium  which is markedly
 under-abundant ([Al/Fe] = $-$0.7).
 Closer study shows that sodium may be  overabundant ([Na/Fe] = +0.3) and
 a suite of
 elements are mildly under-abundant: Ca, Sc, Ti, Zr and Ba with [El/Fe]
 $\simeq -0.35$.

 This abundance pattern is
  largely reminiscent of the
 pattern
 exhibited by stars affected by dust-gas separation, i.e., the
 photosphere is
dominated by accretion of gas but not dust from a cool envelope, possibly
 a circumbinary
 dusty disk. Dust-gas separation is most remarkably exhibited by post-AGB
 binaries such as
 HR 4049 \citep{Van2003}. In such cases, the abundance
anomalies
 are correlated with the condensation temperature ($T_c$): elements
 with high $T_c$
 are most under-abundant. In Figure 3, we plot [X] (where [X] =$\rm log\epsilon(X)_{*}$$-$$\rm log\epsilon(X)_\odot$) versus the $T_c$
 computed by Lodders (2003)
 for a solar composition mixture. Elements N and O depart from a
 general tendency for
 [X] to decrease with increasing $T_c$. 
  In presenting Figure 3, the adopted reference has been that
 [X/Fe] $\simeq 0.0$ for
all elements in an unevolved star with [Fe/H] $\simeq -0.4$. This is
 essentially true for a thin disk star, although a dispersion in heavy elements at a given
 [Fe/H] was noted by \citet{Edvar93}.  If SAO 40039 is a thick disk star, the
 $\alpha$-elements in
 Figure 3 should be plotted with [X] about $-0.2$ dex less than shown in the figure
 but the overall
 trend and the scatter about that trend will be unaffected.
The Na, S, and Zn abundances suggest that the initial metallicity of the
star was $\rm [Fe/H]_o$ $\simeq -0.2$.

 The high N abundance is, probably, a consequence of extensive burning of H
 to He by the
 CN-cycle with concomitant conversion of C to N. Survival of C at about its
initial
 abundance suggests that N synthesis occurred primarily after the
 atmosphere was
 enriched with C at the third dredge-up.  One possibility is H-consumption
 during a final
 He-shell flash \citep{Herwig2001,Blocker2001}. The O is
overabundant with
 respect to the
 initial abundance, possibly it was enriched by the third
 dredge-up
 and temperatures during the late episode of H-burning were too low to
 destroy O or non-LTE line formation can reduce the C, N and O
 abundances.
 For an effective temperature of about 8000K,
 \citet{Venn95} give non-LTE correction to C and N abundances from
 C\,{\sc i} and N\,{\sc i}
 lines of about $-0.4$ dex for normal supergiants.
 \citet{Tak98} have estimated NLTE corrections for O\,{\sc i} to be $-$0.1
 dex at
  this effective temperature for supergiants.
 After applying these non-LTE corrections, the above sketch of a possible
 origin
 for SAO 40039 is little affected.

 \section{Concluding remarks}

 Though, the existence of extremely H-deficient stars with He/H ratio of
 about 10,000,
 showing He\,{\sc i} lines in their absorption spectra, is known
 \citep{Pan2001,Pand2006},
 SAO\,40039, from our estimated He/H ratio of about unity, appears to be
 the first detection
 of a mildly H-deficient star.
 \citet{Lam96} had commented on the possibilities of such stars and 
 on the difficulty in detection for stars too cool to show
 He\,{\sc i} lines.
 With an effective temperature of about 8000K, SAO\,40039 seems to be
 responding to this call.

 It is very likely that other post-AGB stars
 may belong to the mildly H-deficient class. 
 Several examples
 like SAO 40039 have detectable He\,{\sc i} lines. The question then arises;
 are post-AGB stars
 of a temperature too low to provide He\,{\sc i} lines also He-rich and how
 can their
 H-deficiency be unmasked? Exploration of post-AGB stars showing He\,{\sc
 i}
 lines need to be continued with models
 of appropriate H/He ratios. Additionally, non-LTE abundance analyses as
 described by \citet{Pandey2011} should be performed
 to further refine knowledge of the chemical compositions.

 We are thankful to the VBO observing support team, in particular to Mr G. Selvakumar, 
for obtaining a spectrum
 of SAO 40039 using the VBT echelle spectrometer.
 DLL thanks the Robert A. Welch Foundation of Houston, Texas for support
 through grant F-634.

 \begin{figure}
 \epsscale{1.0}
 \plotone{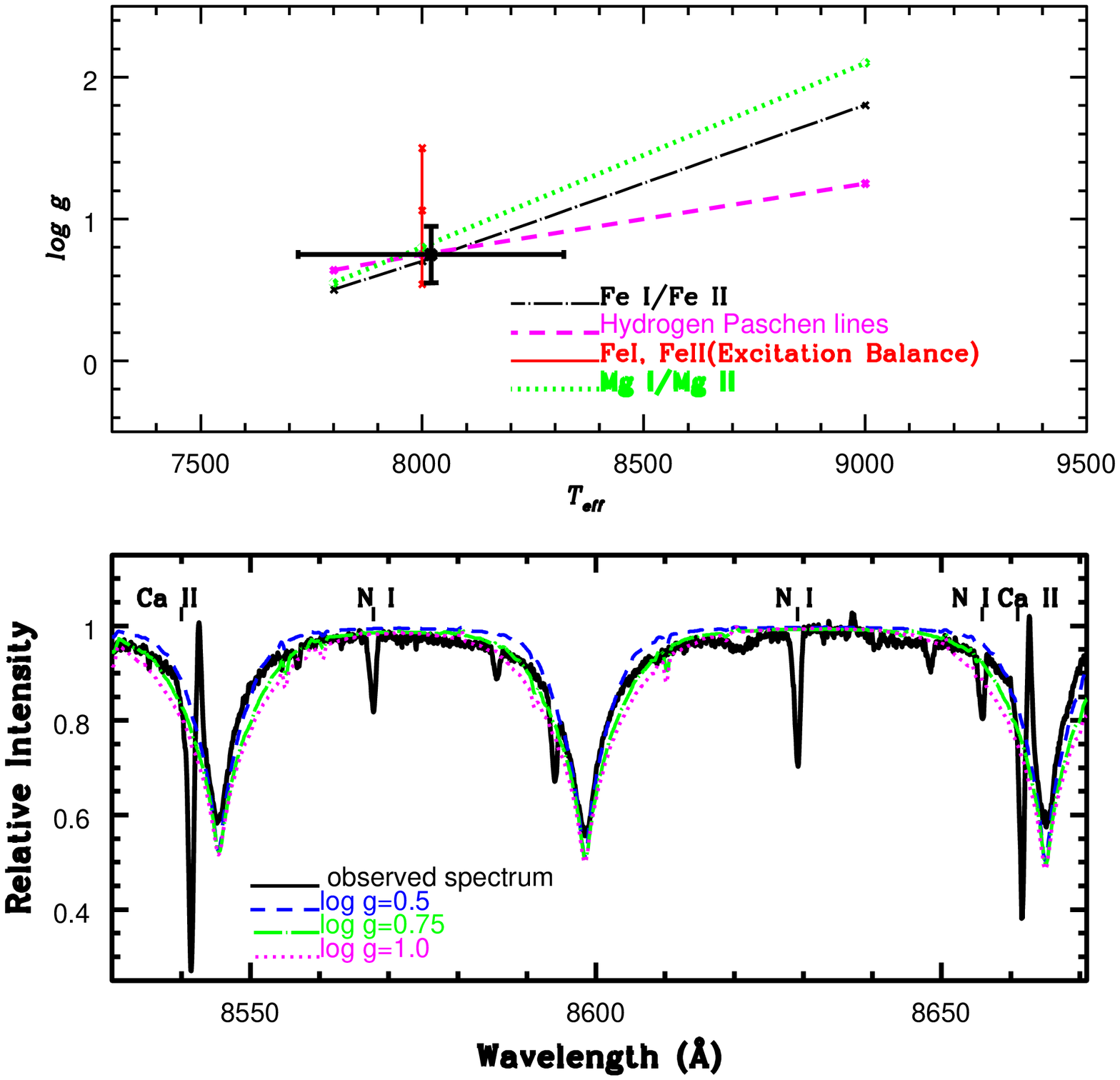}
 \caption{The upper panel shows the loci satisfying the ionization
 balance $-$ see keys on the figure. The locus satisfying the
 H\,{\sc i} Paschen lines is shown by dashed line.
 The solid line represents the excitation balance of both Fe\,{\sc i} and
 Fe\,{\sc ii}. The cross shows the
 derived stellar parameters. 
 Lower Panel shows the observed and synthesized profiles of Paschen lines of  H\,{\sc i}
 in the
  8600\AA\ region for {\itshape T$_{\rm eff}$} = 8000K,
  He/H = 1.0 and {\itshape $\log g$} = 0.5, 0.75 and 1.0 $-$ see keys
  on the figure.} 
 \end{figure}

 \begin{figure}
 \epsscale{1.0}
 \plotone{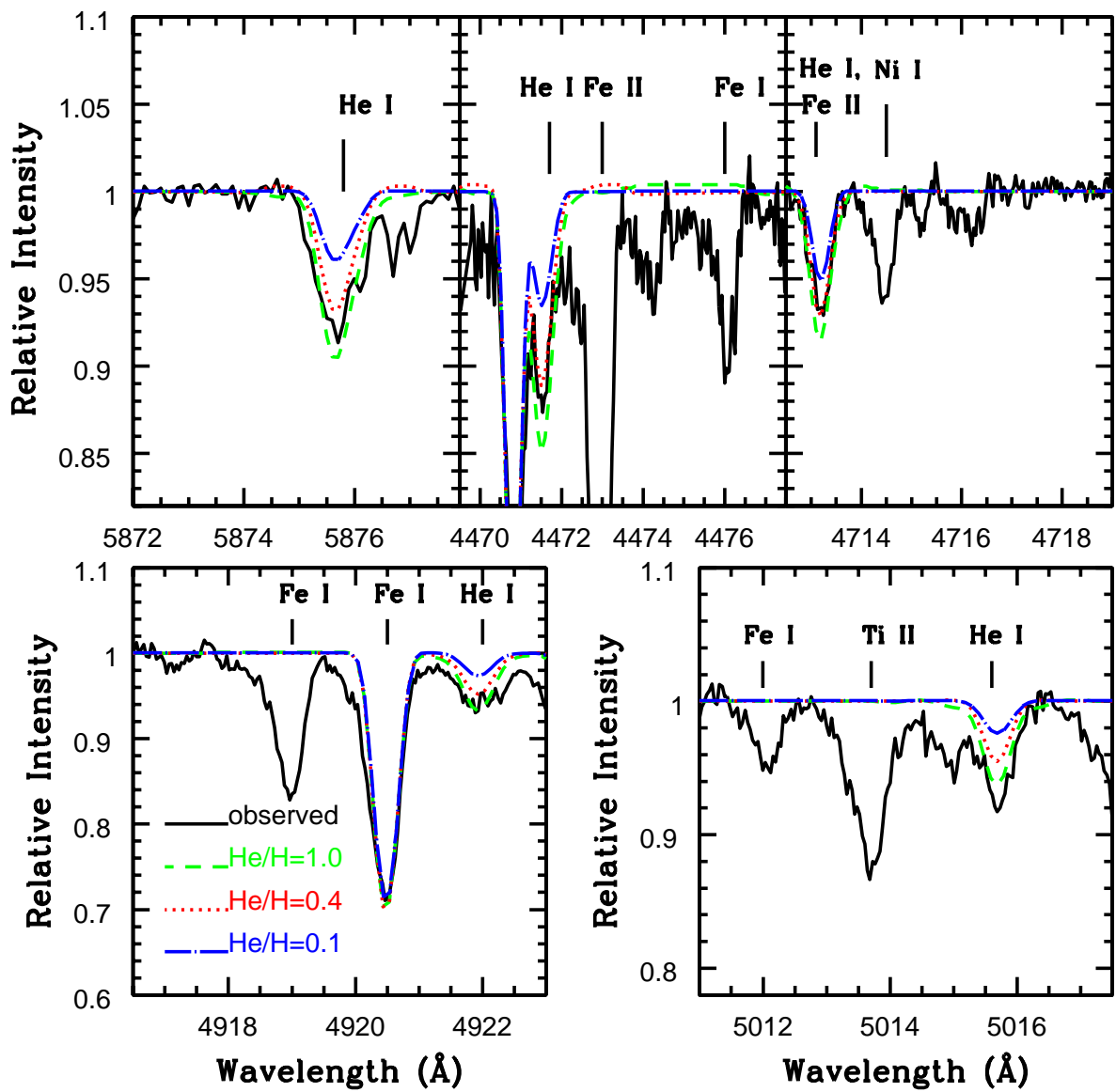}
 \caption{The observed and synthesized He\,{\sc i} profiles for the triplet lines at 5876, 4471 and 4713\AA\ and for the singlet lines at 4922 
 and 5015\AA\ for SAO\,40039 using 
 models with{\itshape T$_{\rm eff}$} = 8000K, {\itshape $\log g$} = 0.75 for
 He/H = 1.0, 0.4 and 0.1 $-$ see keys on the figure.}
 \end{figure}

\begin{figure}
\epsscale{1.0}
\plotone{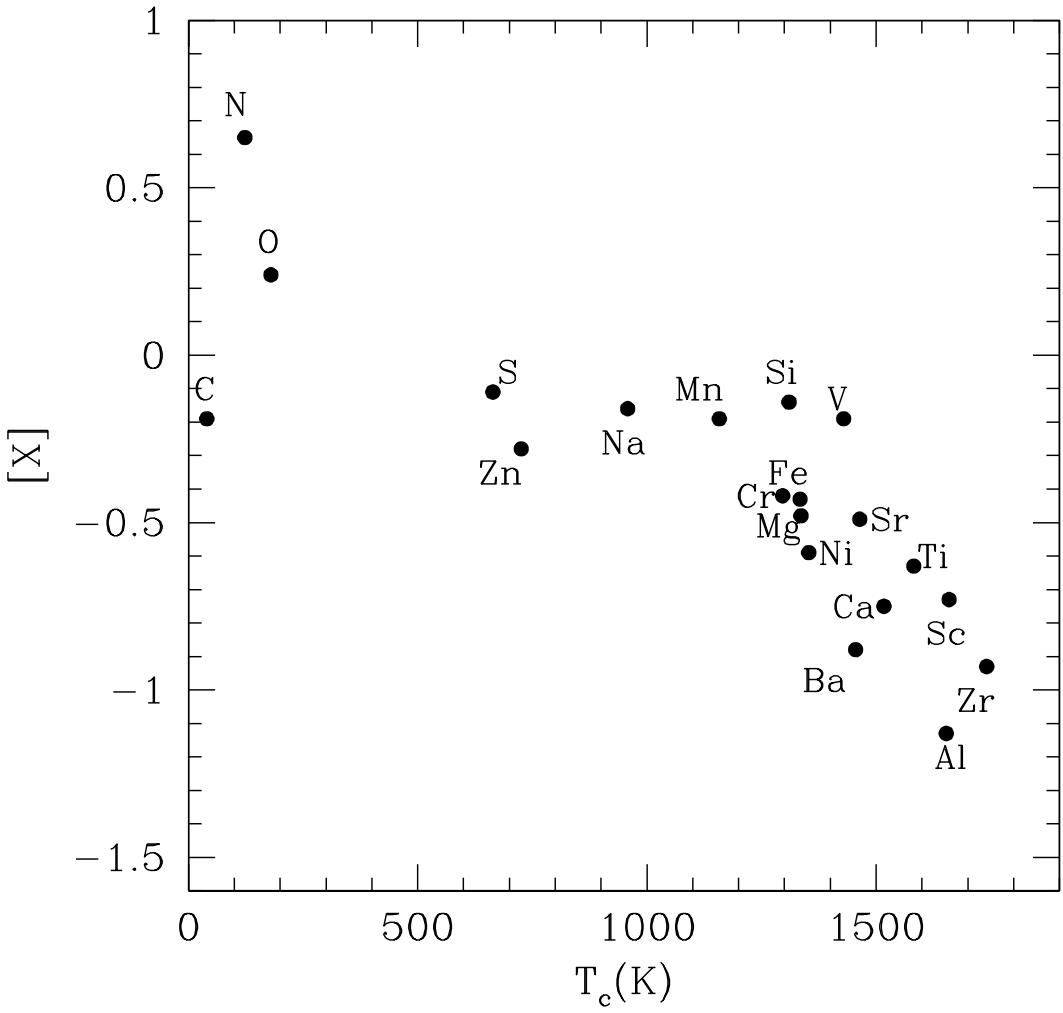}
\caption{The [X] versus $T_c$ plot for SAO 40039.}
\end{figure}

\tabletypesize{\scriptsize}
\begin{deluxetable}{lcccccccccccc}
\tablecaption{Comparison between abundances estimated with MOOG using He/H=0.1 model and the abundances computed with SPECTRUM using the hydrogen deficient (He/H=1.0) model}
\tablenum{1}
\tablehead{
 & & \multicolumn{3}{c}{MOOG\tablenotemark{a}}  &  \multicolumn{3}{c}{SPECTRUM\tablenotemark{b}} &  &  \multicolumn{3}{c}{SPECTRUM\tablenotemark{c}} &  \\

\cline{3-5}  \cline{7-9} \cline{11-12} \\
\colhead{Species} &\colhead{$\log \epsilon_{\odot}$}& \colhead{$\log\epsilon$} &
 \colhead{[X/Fe]}  &\colhead{n} &&\colhead{$\log\epsilon$} & \colhead{[X/Fe]} & \colhead{n} &\colhead{$\triangle$\tablenotemark{d}}
 & \colhead{[X/Fe]}  & \colhead{n} \\
}

\startdata
H I& 12.00 & 12.00 & ...& ...&& 11.45& ...& ...& ...& ...& ... \\
He I& 10.98 & 10.98& ...& ...&& 11.45& ...& ...& 0.2& ...& ... \\
Li I& 3.26& 4.03 &$<$$+$0.68$\pm$0.00& 1&& 3.50&$<$$+$0.67$\pm$0.00 & 1& ...& ... \\
C I& 8.43 & 8.78& $+$0.26$\pm$0.16& 21 && 8.24&$+$0.24$\pm$0.16& 21& 0.3& $+$0.27$\pm$0.18& 9 \\
N I& 7.83 & 9.05& $+$1.13$\pm$0.25& 9&& 8.48& $+$1.08$\pm$0.20& 9& 0.1& $+$1.32$\pm$0.04& 2 \\
O I& 8.69 & 9.47&$+$0.69$\pm$0.15& 8&& 8.93& $+$0.67$\pm$0.14&8& 0.1& $+$0.76$\pm$0.14& 2 \\
Na I& 6.24 & 6.58& $+$0.25$\pm$0.10& 2&&6.08& $+$0.27$\pm$0.07&2& 0.3& ...& ...  \\
Mg I & 7.60 & 7.76& $+$0.07$\pm$0.31& 4&& 7.13& $-$0.04$\pm$0.18&4& 0.4& $-$0.26$\pm$0.11& 2 \\
Mg II& 7.60& 7.78& $+$0.09$\pm$0.15& 4&& 7.12& $-$0.05$\pm$0.15& 4& 0.1& $-$0.10$\pm$0.03& 3 \\
Al I& 6.45 & 5.95& $-$0.59$\pm$0.01& 2 && 5.32& $-$0.70$\pm$0.03& 2& 0.4& ...& ...  \\
Si II& 7.51& 7.90& $+$0.30$\pm$0.16& 3&& 7.37& $+$0.29$\pm$0.13& 3& 0.2& $+$0.31$\pm$0.00& 1 \\
S I& 7.12 & 7.53& $+$0.32$\pm$0.01& 2&& 7.01& $+$0.32$\pm$0.02& 2& 0.3& ...& ... \\
Ca I& 6.34 & 6.18& $-$0.25$\pm$0.12& 2&& 5.55& $-$0.36$\pm$0.16& 2& ...& ...& ... \\
Ca II& 6.34& 6.21& $-$0.22$\pm$0.00& 1&& 5.64& $-$0.27$\pm$0.00& 1& 0.4& $-$0.36$\pm$0.00& 1 \\
Sc II& 3.15 & 2.89& $-$0.35$\pm$0.17&5&& 2.42& $-$0.30$\pm$0.15& 5& 0.3& $-$0.05$\pm$0.16& 2 \\
Ti II& 4.95 & 4.83& $-$0.21$\pm$0.23& 20&& 4.32& $-$0.20$\pm$0.20& 20& 0.3& $-$0.24$\pm$0.16& 12 \\
V II& 3.93& 4.21& $+$0.19$\pm$0.04& 2&& 3.74& $+$0.24$\pm$0.03& 2& 0.2& ...& ...  \\
Cr I& 5.64& 5.62& $-$0.11$\pm$0.13& 3&& 5.11& $-$0.10$\pm$0.10& 3& 0.4 & $-$0.06$\pm$0.23& 2\\
Cr II& 5.64& 5.83& $+$0.10$\pm$0.11& 9&& 5.33& $+$0.12$\pm$0.10& 9& 0.2& $+$0.09$\pm$0.06& 4 \\
Mn II& 5.43& 5.73& $+$0.21$\pm$0.00& 1&& 5.24& $+$0.24$\pm$0.00& 1& ...& ... \\
Fe I& 7.50& 7.54& $-$0.05$\pm$0.18& 26&& 7.02& $-$0.05$\pm$0.17& 26& 0.4& $-$0.01$\pm$0.19&7
\\
Fe II& 7.50& 7.65& $+$0.06$\pm$0.19& 20&& 7.13& $+$0.06$\pm$0.17& 20& 0.2& $+$0.01$\pm$0.14& 9 \\
Ni I& 6.22& 6.16& $-$0.15$\pm$0.15& 4&& 5.63& $-$0.16$\pm$0.14& 4& 0.3& ... \\
Zn I& 4.56& 4.81& $+$0.16$\pm$0.30& 2&& 4.28& $+$0.15$\pm$0.23& 2& 0.4 & ... \\
Sr II& 2.87& 3.01& $+$0.05$\pm$0.07& 2&&2.38&  $-$0.06$\pm$0.03& 2& ...& ... \\
Zr II& 2.58& 2.13& $-$0.54$\pm$0.08&2&& 1.65& $-$0.50$\pm$0.04& 2& 0.3& ... \\
Ba II& 2.18& 1.85& $-$0.42$\pm$0.15& 2&& 1.30& $-$0.45$\pm$0.11& 2& ...& $-$0.35$\pm$0.00& 1 \\
Eu II& 0.52& 0.40& $<$$-$0.10$\pm$0.00& 1&& 0.15& $<$$+$0.06$\pm$0.00& 1& ...& ... \\
\enddata
\tablenotetext{a}{MOOG abundances computed using the He/H=0.1 model for the McDonald spectra.}
\tablenotetext{b}{SPECTRUM abundances computed using the He/H=1.0 model for the McDonald spectra.}
\tablenotetext{c}{SPECTRUM abundances computed using the He/H=1.0 model for the VBO spectra.}
\tablenotetext{d}{$\triangle$ corresponds to the square root of the sum of the squares of the abundance errors due to uncertainties in the stellar parameters: $\triangle${\itshape T$_{\rm eff}$}, $\triangle${\itshape $\log g$} and  $\triangle${$\xi_{t}$}}
\end{deluxetable}
      
\tabletypesize{\normalsize}
\ptlandscape
\hoffset=-0.5in
\begin{deluxetable}{lllll}
\tablewidth{4.2in}
\tablenum{2}
\centering \tablecaption{Detailed line list for SAO 40039\tablenotemark{1}}
\tablehead{
\colhead{Line} & \colhead{$\chi$} & \colhead{\itshape$\log$gf} &\colhead{$W_{\lambda}$}
 & \colhead{$\log{\epsilon}$} \\
 &\colhead{(eV)} &&\colhead{(m$\AA\))}
}
\startdata
He\,{\sc i} $\lambda$5875.61& 20.96& $+$0.739\tablenotemark{a}& Synth& 11.45 \\
He\,{\sc i} $\lambda$4471.47& 20.96& $+$0.053\tablenotemark{a}& Synth& 11.40 \\
He\,{\sc i}$\lambda$4921.93& 21.21& $-$0.435\tablenotemark{a}& Synth& 11.45 \\
He\,{\sc i}$\lambda$5015.67& 20.61& $-$0.819\tablenotemark{a}& Synth& 11.50 \\
He\,{\sc i}$\lambda$4713.13& 20.96& $-$0.975\tablenotemark{a}& Synth& 11.40 \\
C\,{\sc i} $\lambda$4775.87& 7.49& $-$2.304\tablenotemark{a}& 35& 8.41  \\
C\,{\sc i} $\lambda$5039.05& 7.48& $-$2.100\tablenotemark{a}& 33& 8.15   \\
C\,{\sc i} $\lambda$4770.02& 7.48& $-$2.437\tablenotemark{a}& 30& 8.45 \\
C\,{\sc i} $\lambda$4766.67& 7.48& $-$2.400\tablenotemark{a}& 26& 8.34  \\
C\,{\sc i} $\lambda$4826.80& 7.49& $-$2.140\tablenotemark{a}& 24& 8.04 \\
\enddata
\tablenotetext{a}{NIST database (http://physics.nist.gov/PhysRefData/ASD/lines$\_$form.html)}
\tablenotetext{1}{Table 2 is published in its entirety in the electronic edition of \apj. A portion is shown here for guidance regarding its form and content.} 
\end{deluxetable}

 \end{document}